\documentclass{acm}
\usepackage{algpseudocode}
\usepackage{paralist}
\usepackage{booktabs}
\usepackage{caption}
\usepackage{subfig}
\usepackage{url}

\begin{document}


\title{Modeling emergence of norms in multi-agent systems by applying tipping points ideas}


%
%
%
%

%

\numberofauthors{1}

\author{
\alignauthor
Francisco Lopez\\
\affaddr{School of Computer Science}\\
\affaddr{University of Manchester}\\
\affaddr{Manchester, United Kingdom}\\
\email{lopezfrancisco547@gmail.com}
}

%
\maketitle

\begin{abstract}
	Norms are known to be a major factor determining humans behavior. It's also shown that norms can be quite effective tool for building agent-based societies. Various normative architectures have been proposed for designing normative multi-agent systems (NorMAS). Due to human nature of the concept norms, many of these architectures are built based on theories in social sciences. Tipping point theory, as is briefly discussed in this paper, seems to have a great potential to be used for designing normative architectures. This theory deals with the factors that affect social epidemics that arise in human societies. In this paper, we try to apply the main concepts of this theory to agent-based normative architectures. We show several ways to implement these concepts, and study their effects in an agent-based normative scenario.
\end{abstract}





\section{Introduction}
\label{sec:int}

Human societies are simultaneously frustratingly unchanging and yet susceptible to ``epidemics'' that sweep across the social fabric causing people to adopt previously rare practices.   Tipping point theories attempt to explain the subtle triggers behind these social processes.  In 2000, Malcolm Gladwell~\cite{malcolm2000tipping} produced a popular science book summarizing three key factors which trigger tipping points: 1) scale-free networks (the Law of the Few); 2) effective messaging (the Stickiness Factor) and 3) environmental influences (the Power of Context).
This paper relates tipping point theory to the process of norm emergence in multi-agent systems; we propose that normative agent architectures can serve an excellent computational model for expressing many contagious social phenomena, including tipping points and information cascades.

Social norms are known to be a major factor governing humans' behavior; unbeknownst to us, many of our everyday behaviors are influenced by these implicit standards \cite{beheshti2012extracting}.  Various normative architectures have been proposed for designing normative multi-agent systems (NorMAS) capable of reasoning about norm adoption.  Some of these systems have been grounded in social science theory, but the aim of many architectures is simply to effectively address standard multi-agent system challenges, including agreement formation, coordination and conflict resolution \cite{beheshti2009multi}.  

Despite recent research progress in the area, the complete life-cycle of norms is far from fully understood.  The complex nature of human decision-making makes comprehending the rationale behind social interactions difficult, since people are notoriously bad at self-reporting their motivations \cite{beheshti2014normative}.  The field of agent-based modeling aims to create agents in the image of humans.  These agents typically have cognitively-inspired decision-making components, and are situated in life-like scenarios.  In both standard multi-agent systems and cognitively-inspired models, existing social theories have been employed toward the construction of normative models \cite{beheshti2009predicting}.  Various stages of the norm life-cycle including recognition, adoption, compliance and emergence are often modeled on similar concepts in social sciences. 

This paper proposes a unified model of how norm emergence in networked agent societies can be used to predict the effects of common tipping point triggers.  Previous work on norm emergence in networks has investigated the effects of social network topology in static~\cite{villatoro2009topology,sen2010effects} and also dynamic networks~\cite{savarimuthu2009norm}.  Yu et al.~\cite{yu2013emergence} presented an evaluation of different learning methods on norm emergence in networked systems.  In our work, we simply employ network structures as a medium to apply ideas from tipping point theory relating to the Law of the Few. Therefore, the structure of agents' network is not of interest by itself, other than making it congruent with human social networks.
Our main contribution is showing the role and significance of tipping point principles in normative agent systems, and evaluating the potential impact of this model on NorMAS design.  The next section provides an overview of related work in this area.

\section{Related Work}
\label{sec:rlw}

Hollander and Wu \cite{hollander2011current} refer to three categories of normative studies in the social sciences: 1) social function of norms \cite{beheshti2014normative}, 2) impact of social norms \cite{beheshti2014homan}, 3) mechanisms leading to the emergence and creation of norms \cite{beheshti2014hybrid}.  In the context of social function, norms are often concerned with the \textit{oughtness} and \textit{expectations} of agent behavior; where oughtness refers to the condition where an agent should or should not perform an action, and expectation refers to the behavior that other agents expect to observe from that agent \cite{beheshti2013improving, beheshti2015cognitive}. An example of work belonging to this category is Boella and Torre's architecture containing separate subsystems for counts-as conditionals, conditional obligations, and conditional permissions~\cite{Boella06anarchitecture}.

Within the second category, social impact, norms are considered in terms of cost provided to or imposed on the parties involved in a social interaction \cite{beheshti2014negotiations}. For instance, punishment and sanctions are introduced as two enforcement mechanisms used to achieve the necessary social control required to impose social norms~ \cite{villatoro2011dynamic}. Here they demonstrate a normative agent that can punish and sanction defectors and also dynamically choose the right amount of punishment and sanction to impose \cite{beheshti2013analyzing}.

As noted in \cite{hollander2011current}, the third category is concerned with the \textit{how} of norms more than the \textit{why.} Relevant work in NorMAS domain that falls into this category is divided into the following groups  \cite{beheshti2013agent}.  The first group is composed of research in which norms are practically hard-wired into the system or dictated directly to the agents.  In the second set, norms emerge from social interactions among agents. Sen and Airiau's work~\cite{sen2007emergence} in which agent interactions are modeled using payoff matrices, focuses on norm emergence through social learning in agent societies. Conte et al.'s \textit{EMIL} was designed as an architecture for modeling norm emergence~\cite{conte2013minding}. The EMIL architecture includes a dynamic cognitive model of norm emergence and innovation \cite{beheshti2011new}.

Much of the existing work in normative multi-agent systems explicitly or implicitly relies on social science theories \cite{beheshti2009new}.  In a recent work, some of the well-known theories of philosopher David Hume were evaluated using an agent-based model called HUME$_{2.0}$ \cite{conte2013minding}.  This work demonstrates how social justice concepts can even emerge from heterogeneous agents that are not endowed with norm representations.
 
Self-determination theory is also referenced by some of the normative works \cite{beheshti2009pairwise}.  Here the focus is on the agents' motivation and the extent to which the motivation is intrinsic or extrinsic.  Neumann studied existing normative architectures to see how much they comply with self-determination theory~\cite{neumann2010}. 

Practice theory is an example drawn from anthropology; this theory describes how changes in the society are based on the interactions between the human agents and social structure.  For instance, an agent-based model for energy demand and supply social practices is presented in \cite{balkemodelling}, which shows how energy consumption norms form and evolve in urban societies.  The next section provides background on tipping point theory, primarily from a sociology perspective.

\section{Tipping Point Theory}
\label{sec:tip}
The term, ``tipping point'' was initially coined in physics to describe the situation in which the state of an object rapidly changes from one stable equilibrium to another different equilibrium. Morton Grodzins was the first to use term in social sciences to describe an interesting phenomenon he observed in the US cities, known as \textit{white flight}~\cite{grodzins1957metropolitan}. His observation was that in some metropolitan areas the percentage of African-American people would increase up to a certain point. After that point, those with white ethnicity immigrated from those cities in large numbers.  Thomas Schelling presented \textit{the general theory of tipping}, which describes how individuals' micromotives and microbehavior can aggregate in the big picture \cite{schelling2006micromotives}. The model of collective behavior introduced by Mark Granovetter~\cite{granovetter1978} uses thresholds to determine the path of social events. This model was initially used to describe how fads are created. 

In normative studies, tipping points are usually denoted as the point of maximum return at which time the behavior has the highest level of acceptability from the population. For instance, in a certain group of friends, the number of times they shower in a week may vary, but a specific value has the highest acceptability by group members as the conventional pattern of behavior.  In this paper, we study the impact of Gladwell's three factors on norm emergence in agent-based normative systems and demonstrate practical ways to apply this versatile theory.

\section{Experimental Setup}
\label{sec:exp}
For our experiments, we employ the classic scenario, rules of the road, that is frequently used to study normative behavior in multi-agent systems. In this scenario, there exists a population of agents that do not have any preference toward driving on the left or right side of a two-way road. No rules or higher enforcement exist to determine the preferred side. This scenario represents a two-action stage game that models the situation where agents need to agree on one of several equally desirable alternatives. The societal norms that we would like to evolve are either driving on the left or driving on the right \cite{sen2007emergence}.

 In this scenario agents receive a fixed value reward and punishment based on the following payoff matrix shown in Table~\ref{tbl:poff}.

\begin{table} [htb]
  \centering
  \includegraphics[width=0.14\textwidth]{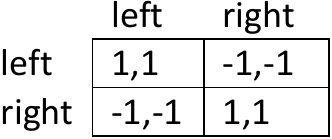}
  \caption{Payoff matrix for rules of the road scenario}
  \label{tbl:poff}
\end{table}

As Yu et al.~\cite{yu2013emergence} note, although this payoff matrix appears simple, the coordination game poses a very challenging puzzle for human beings to solve efficiently. The game has two pure Nash-equilibria: both agents drive left or both agents drive right. Classical game theory, however, does not give a coherent account of how people would play a game like this. The conundrum is that there is nothing in the structure of the game itself that allows the players (even purely rational players) to infer what they ought to do. In reality, people can play such games because they can rely on some contextual cues to agree on a particular equilibrium~\cite{young1996economics}.

In similar studies on normative systems, usually the cumulative payoff (reward) of the whole population of agents is used as a measure of comparing various methods (see \cite{sen2007emergence} and \cite{yu2013emergence} for examples). Instead, we opt to use the norm emergence time for each method as an evaluation metric.  This is functionally equivalent since the payoff received by all agents post norm emergence is the same, hence a method which leads to faster norm emergence will also yield the higher cumulative payoff.
\section{Key few members}
\label{sec:kfm}
In this section, we study the effects of key members of an agent society on the rate of norm emergence.  These key members are selected using standard heuristics for measuring influence within a network; we evaluate the performance of three centrality measures: degree, closeness, and betweenness.  Degree centrality measures the number of edges connected to a node.  Closeness is calculated based on the total distance to all other nodes. Nodes with a high betweenness centrality fall on a large proportion of the shortest paths (geodesics) in the graph.

To model the characteristics of a real social network, we use an algorithm introduced in \cite{wang2011leveraging} to create a synthetic network which follows power law degree distribution and exhibits homophily, a greater number of link connections between similar nodes.\footnote{Commonly described as ``birds of a feather flock together''~\cite{McPherson2001}}  The network generator uses link density (\textit{ld}) and homophily (\textit{dh}) to govern network formation.  A simplified version of the pseudo-code for this method is shown in Figure~\ref{fig:ps}.  For our model, we assumed predefined values for \textit{ld} and \textit{dh}. The nodes of the graph represent the individuals (agents) in the simulation, who can be considered as car drivers.

\begin{figure}[h!t]
\begin{tabular}{p{0.9\columnwidth}}
\\
\toprule
\end{tabular}
\begin{algorithmic}
	\State $G = \text{Null}$
 \Repeat 
 \State sample $r$ from uniform distribution $U(0,1)$
 \If{$r  \medskip \leq  \medskip ld$} 
  \State randomChooseSource($G$)
  \State determineCandidateSink($dh$,$G$)
  \State pickSink()  \Comment \textit{based on power-law distribution}
  \State connect(source,sink)
\Else
\State add a new node to G
\EndIf
 \Until desired number of nodes added to the network
\end{algorithmic}
\begin{tabular}{p{0.9\columnwidth}}
\bottomrule\\
\end{tabular}
\caption{Synthetic friendship network generator}\label{fig:ps}
\end{figure}

We use a weighted voting approach (also known as a structure based method) to determine an agent's decision with regard to its neighbors. The weight for each of an agent's neighbors is computed using a normalized value of that neighbor's centrality value as shown in Equation~\ref{eqn:weight}.

\begin{equation}
	\text{weight}_{i,j} = \dfrac{C_j}{\Sigma_{k=1}^{Deg_i}C_k }
\label{eqn:weight}
\end{equation}

 This equation shows the weight of the link connecting neighbor $j$ to node $i$. $C$ refers to the corresponding centrality value (degree, betweenness and closeness). Also, $Deg_i$ denotes the number of neighbors for node $i$.   The top 10 percent of the population of agents with the greatest centrality values are assumed to be the key elements of a society. At the beginning of our experiments, all of the agents follow a single norm; in other words, all of them have learned (through social learning \cite{sen2007emergence}) to always drive  on one side of the road. In our implementation, each agent has a utility value defined for each of four possible cases: Up-Left, Up-Right, Down-Left and Down-Right, where Up and Down determine the section of road, and Left and Right determine the direction an agent drives. These values are updated while receiving payoffs based on the matrix shown in Table~\ref{tbl:poff}. 

\begin{figure}[htb]
  \centering
  \includegraphics[width=0.47\textwidth]{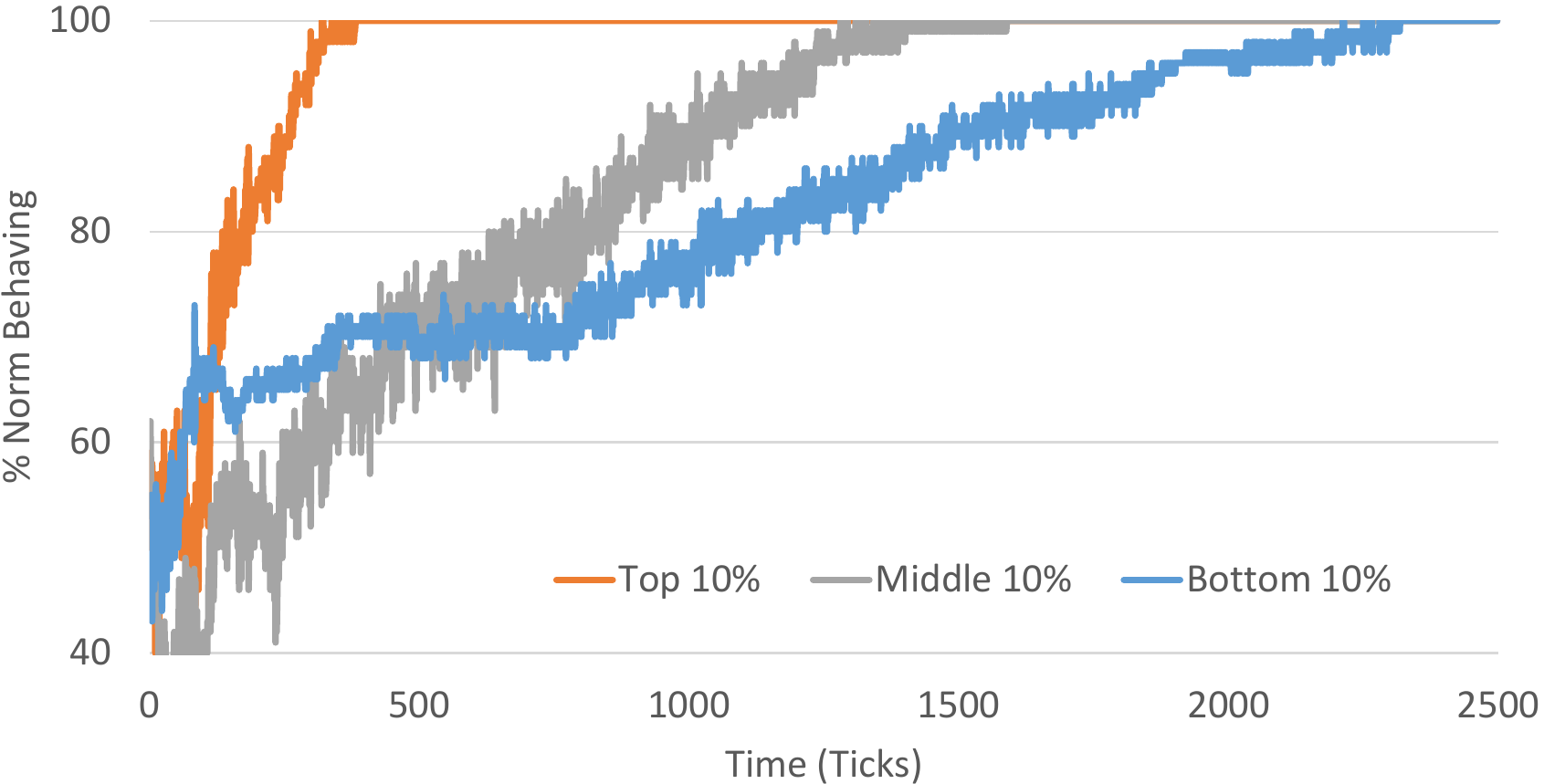}
  \caption{Average number of iterations until the emergence of one norm, when using degree centrality to determine key agents.}
  \label{fig:dgr}
\end{figure}

\begin{figure}[htb]
  \centering
  \includegraphics[width=0.47\textwidth]{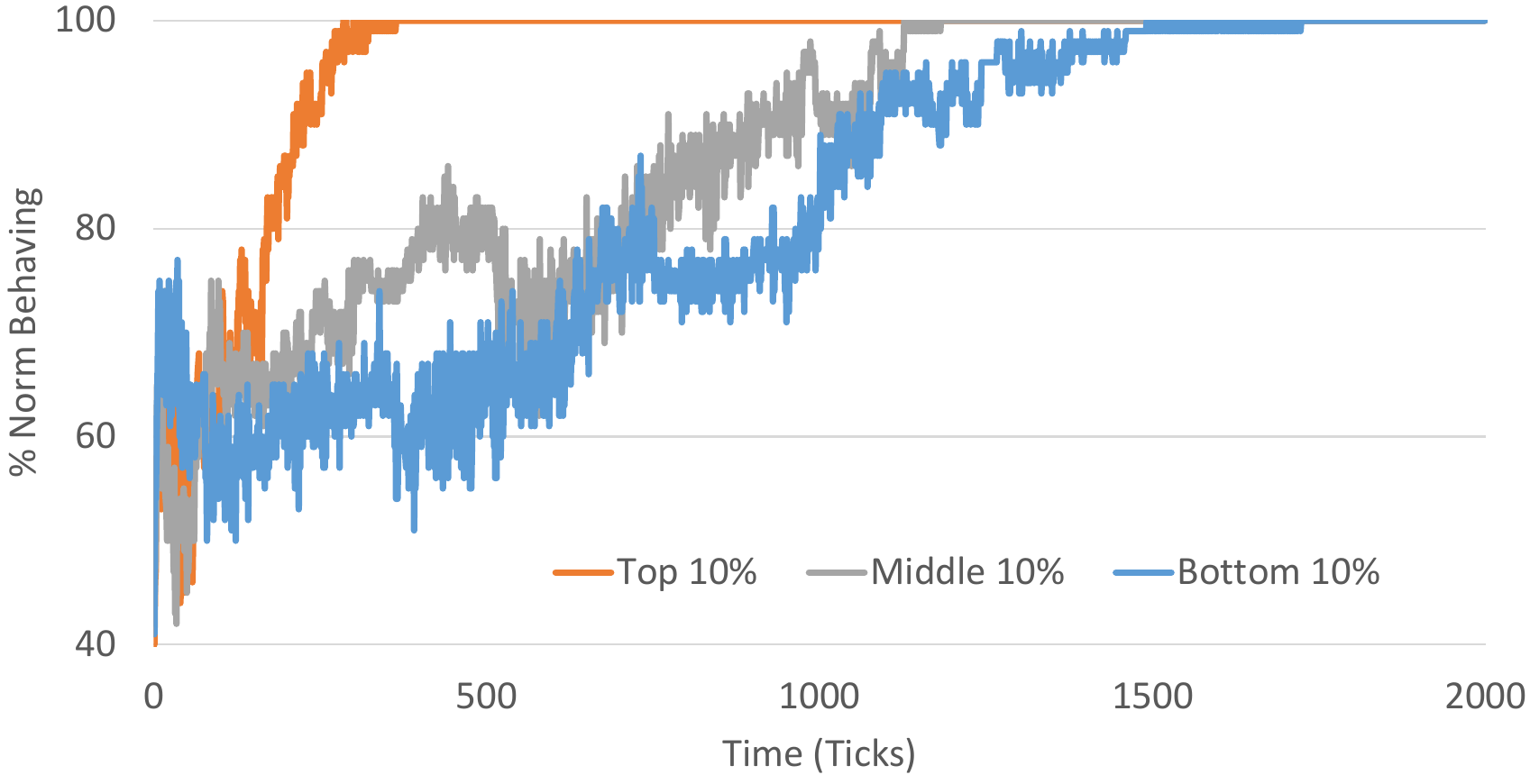}
  \caption{Average number of iterations until the emergence of one norm, when using betweenness centrality to determine key agents.}
  \label{fig:btw}
\end{figure}

\begin{figure}[htb]
  \centering
  \includegraphics[width=0.47\textwidth]{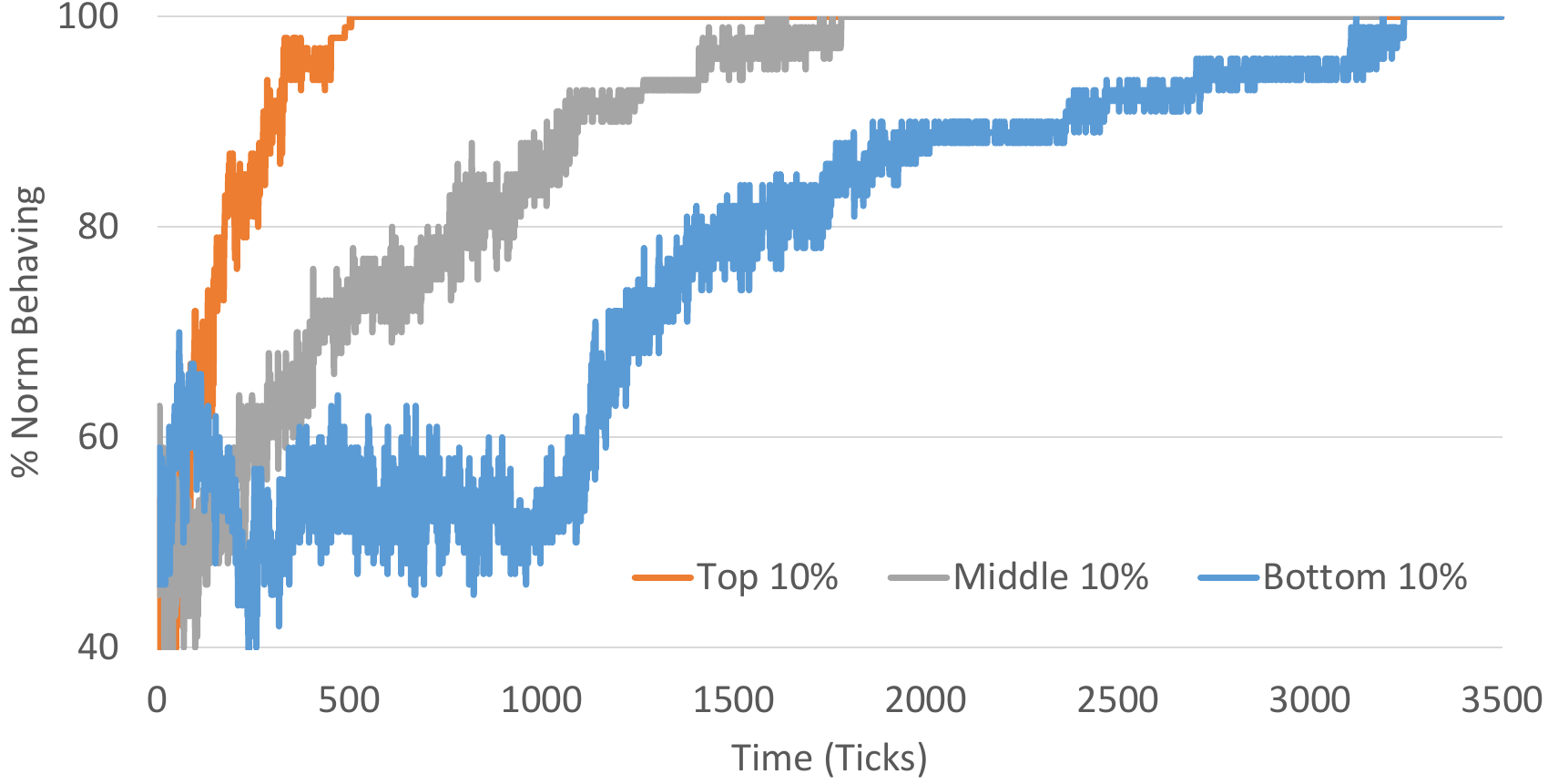}
  \caption{Average number of iterations until the emergence of one norm, when using  closeness centrality to determine key agents.}
  \label{fig:cls}
\end{figure}

In our experiments, we compare the penetration of norm changing behaviors that emanate from key members of a society vs.\ other cases.  
We compare emanation from the top to emanation from the middle and bottom 10 percent of the population. At the beginning of the simulation, the agents (nodes) are ranked based on their centrality value to determine the top, middle and bottom agents.  The utility value of these agents is kept fixed. Neighbors of these agents continue updating their behavior until a new norm emerges in the system. Figure \ref{fig:dgr}, Figure \ref{fig:btw}, and Figure \ref{fig:cls} show the number of iterations required for each case to converge. The population of agents contained 100 agents, and the reported results show the average values over 20 runs. 

The pattern observed in all of three cases was very similar.  When the norm propagation starts from the top 10\% of the population, the norm emerges much faster compared to the other cases. Moreover, there is a fairly sizable difference among top, middle and bottom agents.
The magnitude of difference between the top  and middle 10\% is more than the difference between the middle and bottom. These results are consistent with the role of connectors in tipping point theory.

\newpage
\section{Stickiness Factor}
\label{sec:stf}
According to the tipping point theory, the extent and rate of emerging social norms in a society is not only related to the members of the society, but also related to the content of the message.  An effective message needs to be interesting or ``sticky'' enough to remain in agents' minds. This factor is almost completely independent of the society and its structure, and is a property of the idea.

As Gladwell~\cite{gladwell2006tipping} points out, it is potentially very complicated to determine if a certain message has the necessary stickiness or not, but one characteristic that is usually common to sticky ideas is that it frequently returns to a person's mind. This could be in the form of a desire to sit and watch a popular TV show every night, or in a more extreme case, a clinical addiction to smoking or gambling. Conventional marketing and advertising domains refer to this phenomenon as \textit{rule of 27}. According to this rule, a message (advertisement) should be seen at least 27 times, if the message is going to stick~\cite{pepper2008start}.

\begin{figure}[htb]
  \centering
  \includegraphics[width=0.47\textwidth]{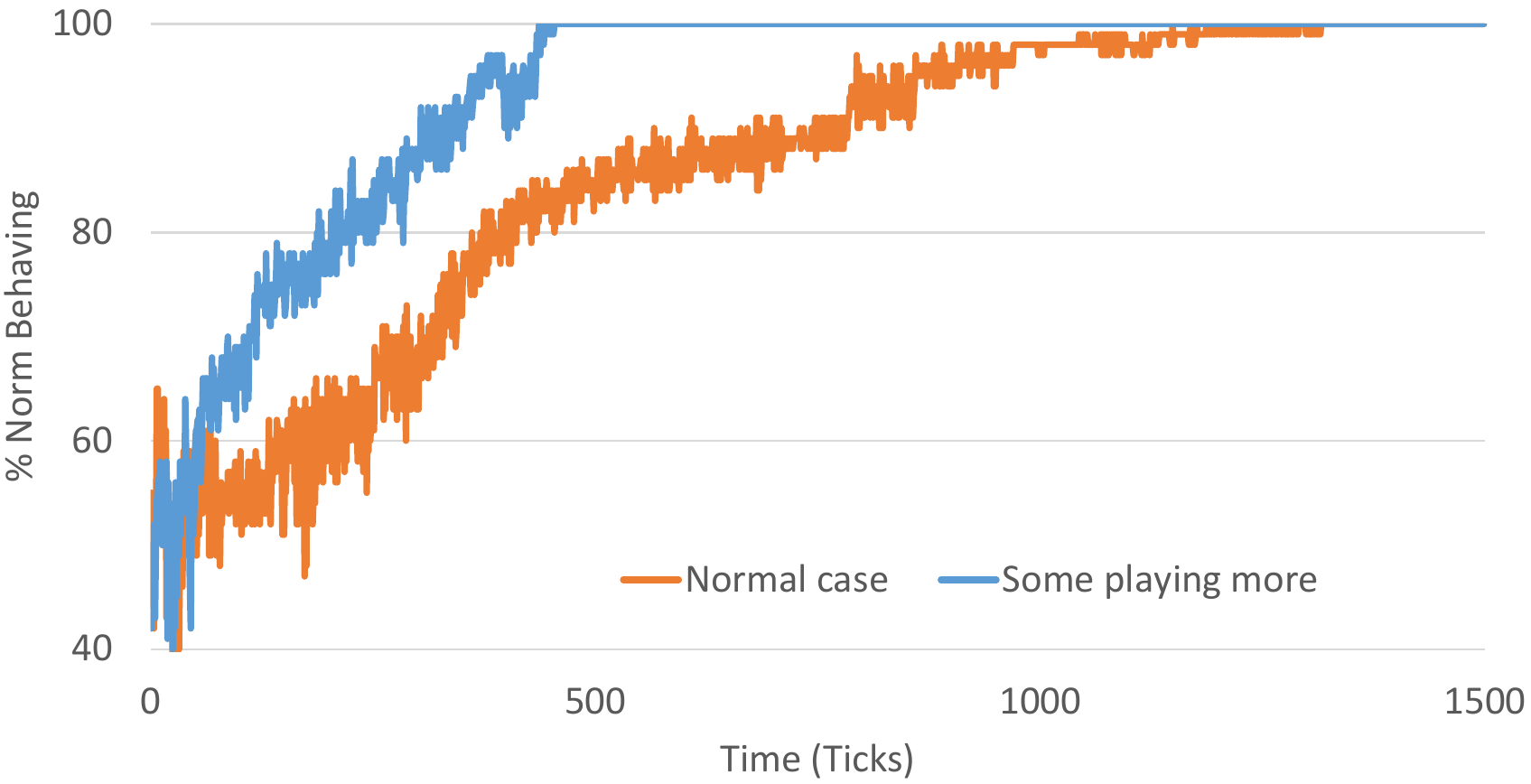}
  \caption{Average number of iterations until the emergence of one norm, when 2 out of 4 agents with fixed utility values play twice with each agent that they encounter.}
  \label{fig:gmr}
\end{figure}

\begin{figure}[htb]
  \centering
  \includegraphics[width=0.47\textwidth]{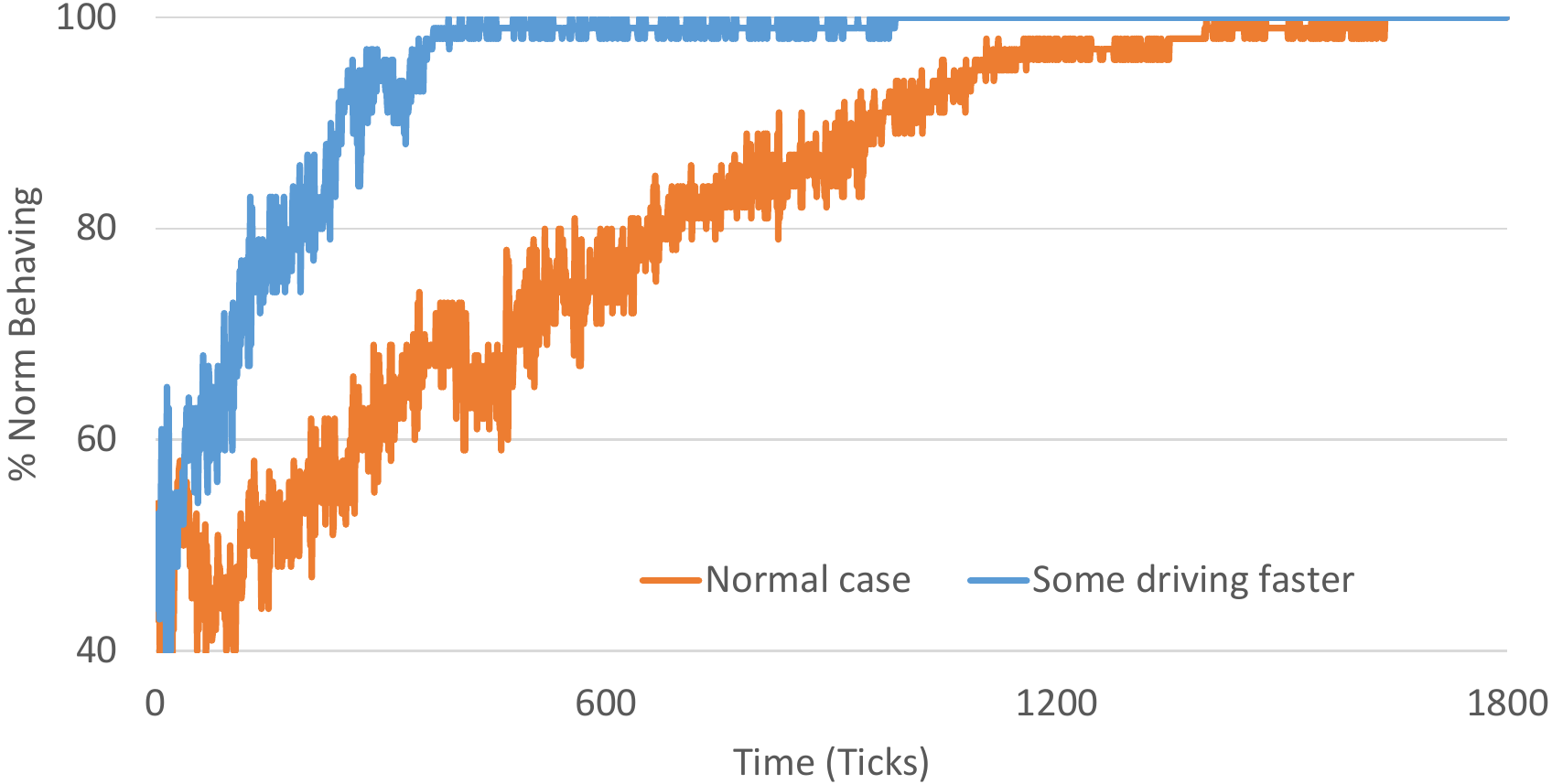}
  \caption{Average number of iterations until the emergence of one norm, when 2 out of 4 agents with fixed utility values go (drive) faster.}
  \label{fig:fst}
\end{figure}

In order to model this property, we assume that the stickiness is represented by the number of games that an agent plays with another agent. Therefore a higher number of games will result in the same effect as a stickier belief.  In our experiments, this idea is evaluated in two different ways. The first way is to increase the number of games that a certain set of agents play. The second way is to have a certain number of agents driving faster than other agents to be exposed to more cars.

Figures \ref{fig:gmr} and \ref{fig:fst} show results related to these two cases. In both cases, we have our original 100 agents plus a group of 2 agents which have a fixed preference to drive on either the left or right. In the first scenario, one group of agents plays two games each time it encounters another agent. In the second scenario, one group of agents moves faster. Both of these scenarios lead to the same effect: increasing the number of times that an agent is exposed to an idea.  This simulates the property of frequently returning to a person's mind. In both cases, when the stickiness factor is implemented, the entire system converges to a single norm faster.

\section{Power of Context}
\label{sec:pcon}
The third element of the tipping point theory refers to the power of context. As Gladwell points out: it is possible to be a better person on a clean street or in a clean subway, than in one littered with trash and graffiti \cite{gladwell2006tipping}. The idea is mostly based on what known in criminology as the theory of  \textit{broken windows}~\cite{wilson1982broken}. According to this theory, slight changes in the environment could result in tipping effects over the whole society. 

\begin{figure}
  \centering
  \includegraphics[width=0.47\textwidth]{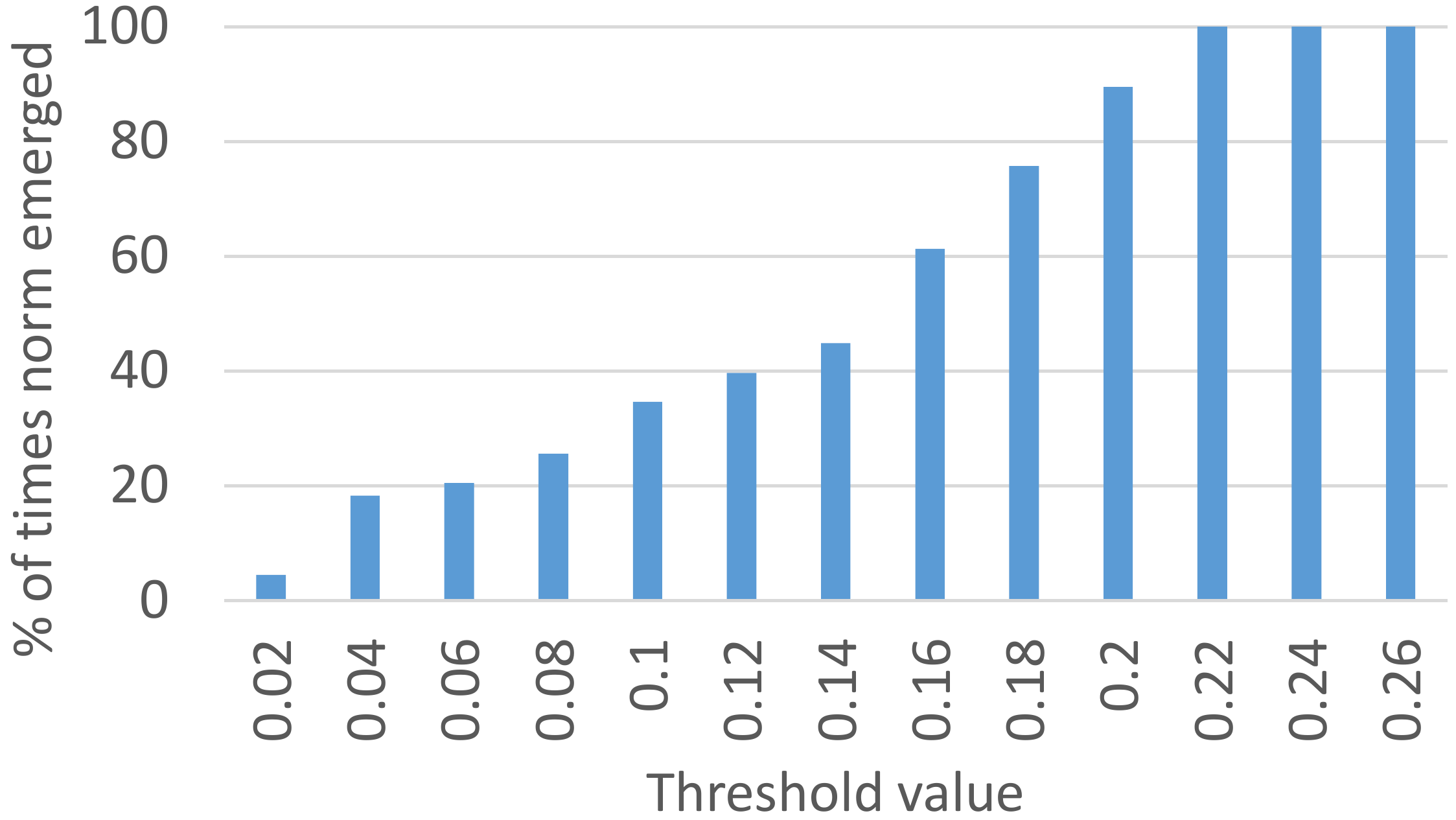}
  \caption{Percentage of times that a norm emerges in the population, when agents have different threshold values for activating.}
  \label{fig:env}
\end{figure}

In order to apply this part of the tipping point theory, we use ideas from the methods for studying fads and cascading effects in networks \cite{watts2001simple}. First, we build a network using the same approach described in Section \ref{sec:kfm}. Then, we assign a threshold value for each agent. Similar to the probabilistic information cascade models, if the cumulative value of the perceived cascade is less than the threshold, nothing will change. If it's higher, the agent will change its current behavior, which in our scenario would result in driving on the other side of the road. Figure \ref{fig:env} shows the percentage of times that a norm emerged in the system for a set of threshold values. The columns show the average results over 20 runs. Agents were selected randomly as a source of a small initial shock in the network, which results in negating the current payoff values for driving on each side of the road.  The frequency of shocks is determined randomly. The system runs until it reaches some fixed iteration number (50,000), unless a different norm is observed.  This experiment illustrates how minor shocks can shape a population fad, resulting in a population-level behavior change. The shocks (pulses) in this model can be viewed as any of the small changes that tipping point theory predicts can result in large changes in the whole society. According to the results presented in Figure \ref{fig:env}, thresholds as small as 0.02 can lead to the emergence of norms in the system in almost 5 percent of the experiments. The computed values for each agent are compared to its tipping point value (normalized between 0 and 1).

There is a second aspect to the power of context, which refers to the number of people in groups. The Rule of 150 says that the size of groups is a subtle contextual factor that makes a big difference. This number is referred as \textit{Dunbar's number} \cite{dunbar1992neocortex}, after the anthropologist who originally proposed the idea.  In groups with fewer than 150 members, people will cooperate relatively easily and rapidly become infected with the community ethos.  Once that threshold is crossed people begin to behave very differently. 150 is our \textit{social channel capacity} as determined on the basis of personal loyalties and 1-on-1 contacts. Beyond the tipping point of 150 the group dynamics simply become too complex. For the average person there are just too many relationships to manage. The group then becomes divided and alienated, and usually splits into two.  Smaller groups have been shown to be more effective at tasks than larger groups. This may be due to biological limitations of humans which make it very difficult for them to handle a larger community.

With the growth of virtual social media sites and the spread of online groups, there has been renewed interest in evaluating the importance of this limit on Facebook~\cite{Economist2009}, Myspace~\cite{gibbons2014modeling} and within massively multiplayer online role-playing games (MMORPGs). The pivotal issue here is that a person cannot maintain a close relationship with all of the members of a larger group which ultimately sabotages its success. Having a direct connection with each member of the group is a necessary component to having a positive social relationship.

	\begin{figure}
  \centering
  \includegraphics[width=0.47\textwidth]{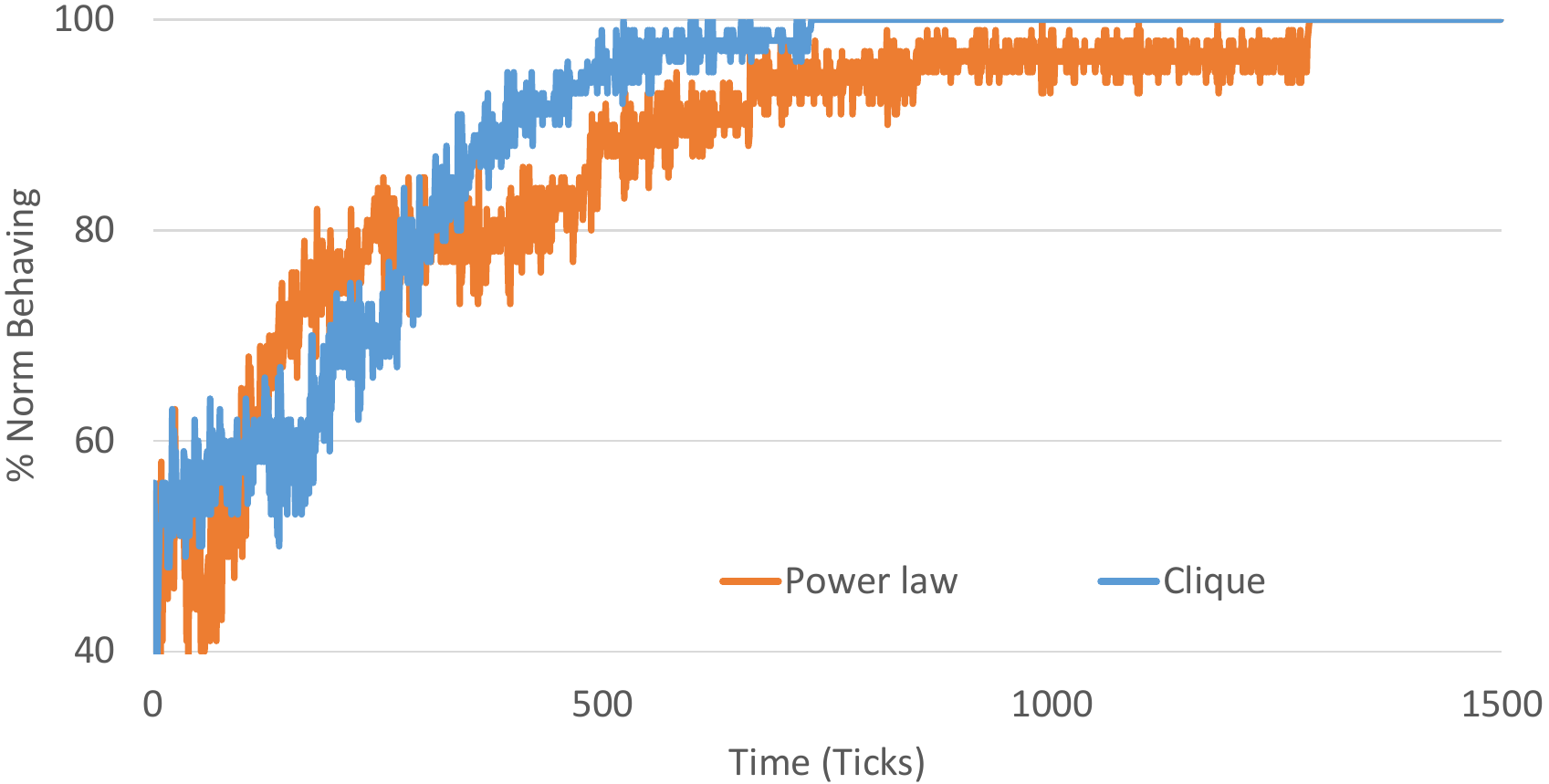}
  \caption{Average number of iterations until the emergence of one norm, when the network structure of agents follows a power-law distribution  and when the network is a complete clique.}
  \label{fig:150}
\end{figure}

We propose using a clique structure to illustrate this idea. In a clique each node has a direct edge to all of other nodes. There are $n*(n-1)$ edges in the resulting graph. We opt to use a directed graph, as that seems to be the general assumption for friendship networks. We compare the emergence of driving norms in a network generated using the method described in Section \ref{sec:kfm}. It should be noted that having more edges does not result in faster convergence. More connections makes the diffusion of ideas easier, while it makes it harder for the agents to find an idea that all agents like. In a clique structure the major voting approach and the weighted voting approach (using the number of edges) are the effectively same, so neither of them elicits earlier norm emergence. Figure \ref{fig:150} shows the number of iterations that were required on average for the two cases to reach norm emergence. The driving norm emerged faster in case of the clique structure than in the power-law degree distribution network.  This shows the potential benefit of such a structure in constructing agent systems, at least for ideal cases.

\section{Conclusion and Future Work}
\label{sec:con}
Norms are complex social behaviors that have been extensively studied in sociology, psychology, and other related fields.  Most normative architectures draw upon theories from the social sciences.  The theory of tipping points has inspired much research in different disciplines. For this paper, we model some of the well-known elements of this theory, as applied to networked agent populations. We illustrate how three of principle ideas including key few members, stickiness factor, and the role of environment can affect the process of norm emergence.  This paper is meant to be an initial step for convincing the NorMAS community of the importance of tipping point theory concepts.   For future work, we are interested in mapping the performance of our normative model to a real-dataset.


\pagebreak 
\bibliographystyle{abbrv}
\bibliography{mine,refs}  
%

\balancecolumns 
\end{document}